\documentclass[manuscript]{aastex}
\usepackage{natbib}
\usepackage{booktabs}
\usepackage{threeparttable}
\begin{document}

\title{On the generation mechanism of electromagnetic cyclotron waves in the solar wind: statistical results from \emph{Wind} observations}

\author{G. Q. Zhao\altaffilmark{1,2}, H. Q. Feng\altaffilmark{1}, D. J. Wu\altaffilmark{3}, G. Pi\altaffilmark{4}, and J. Huang\altaffilmark{2}}
\affil{$^1$Institute of Space Physics, Luoyang Normal University, Luoyang, China}
\affil{$^2$CAS Key Laboratory of Solar Activity, National Astronomical Observatories, Beijing, China}
\affil{$^3$Purple Mountain Observatory, CAS, Nanjing, China}
\affil{$^4$Faculty of Mathematics and Physics, Charles University, Prague, Czech Republic}

\begin{abstract}
Electromagnetic cyclotron waves (ECWs) near the proton cyclotron frequency are frequently observed in the solar wind, yet their generation mechanism is still an open question. Based on the \emph{Wind} data during the years 2005$-$2015, this paper carries out a statistical study on plasma characteristics associated with the occurrence of ECWs.
The probability density distributions (PDDs) of proton temperature anisotropy ($T_\perp/T_\parallel$) and proton parallel beta ($\beta_\parallel$) are investigated, where $\perp$ and $\parallel$ refer to perpendicular and parallel to the background magnetic field, respectively. The PDDs depend on solar wind types as well as wave polarizations, and those for ECWs with left-handed (LH) polarization exhibit considerable differences from the PDDs for ambient solar winds. The distributions of occurrence rates of LH ECWs in ($\beta_\parallel$, $T_\perp/T_\parallel$) space show a tendency that the occurrence rates increase with proton temperature anisotropies. The $\beta_\parallel$ with maximum of occurrence rates is near 0.1 when $T_\perp/T_\parallel > 1$ while it is around 1 when $T_\perp/T_\parallel < 1$. The presence of alpha$-$proton differential flow with large kinetic energy corresponds to a much high occurrence rate as well as the domination of LH polarization of ECWs. Based on these observations and existing theories, we propose that the proton cyclotron and parallel firehose instabilities with effects of alpha$-$proton differential flow are likely responsible for the local generation of LH ECWs in the solar wind. The generation mechanism of right-handed ECWs seems to be complicated and more discussions are needed in future researches. 
\end{abstract}


\section{Introduction}
It is well acknowledged that the solar wind is a highly ionized, magnetized plasma streaming outward from the Sun \citep[e.g.,][]{par58p64,gri60p61,neu62p95,axf85p75,han12p89,abb16p55}. It consists of mainly protons, electrons, and minor alpha particles \citep[e.g.,][]{ogi74p95,mar82p35,kas07p01,wan16p21,fuh18p84}. These particles are generally collisionless and far from thermodynamic equilibrium. Proton and alpha particle populations have usually different velocities \citep{mar82p35,mar91p52,kas08p03}. The proton velocity distribution functions often exhibit a secondary, beam component, streaming with respect to the denser, core component \citep{fel73p17,goo76p50,mar82p52}. Moreover, each population (or component) is frequently characterized by a temperature anisotropy \citep{mar82p30,mat13p71}. These nonthermal ions can serve as free-energy sources to excite kinetic waves \citep[e.g.,][]{hol75p63,sch80p13,tuc95p01,mar04p02,mar06p01,cra14p16,hej15p31,hel16p32,wil18p41,kle18p02}.
For instance, the plasma with proton perpendicular temperature ($T_{\perp}$) larger than parallel temperature ($T_\parallel$) can excite proton cyclotron waves by proton cyclotron instability, while a plasma with a converse temperature anisotropy ($T_{\perp} < T_\parallel$) may generate magnetosonic waves by parallel firehose instability \citep[e.g.,][]{gar93,Gar15p49, yoo17p04}. Theoretically, both kinds of waves have similar properties in the solar wind but inherently different polarization senses. Proton cyclotron waves have property of left-handed (LH) polarization while magnetosonic waves are characterized by right-handed (RH) polarization in the plasma frame.

In recent years, a series of studies showed that electromagnetic cyclotron waves (ECWs) near the proton cyclotron frequency can be widely observed in the solar wind \citep{jia09p05,jia10p15,jia14p23,boa15p10,wic16p06,gar16p30,wei16p53,zha17p79,zha18p15}. These ECWs appear as transverse waves with coherent wave forms and propagate mainly in the directions quasi-parallel (or antiparallel) to the background magnetic field. They can occur sporadically with short durations of a few seconds, or last incessantly for several tens of minutes \citep{jia14p23,boa15p10,zha18p15}. The majority of ECWs have amplitudes less than 1 nT, while some ECWs share a large amplitude comparable to the background magnetic field \citep{zha18p15}. They can be of LH polarization or RH polarization with respect to the background magnetic field; the polarization is described in the spacecraft frame throughout the paper, except that we emphasize it is in the plasma frame. The occurrence rate of LH ECWs is usually larger than that of RH ECWs \citep{jia09p05,jia10p15,jia14p23,boa15p10}. Note that the polarization will reverse in the two different reference frames if these waves propagate toward the Sun. The reverse is due to the presence of large Doppler shift resulting from fast movement of the solar wind relative to the approximately standing spacecraft as well as that the speed of the solar wind is several times greater than the phase velocity of ECWs \citep[][]{jia09p05,gar16p30}.

For the generation of ECWs in the solar wind, two versions have been proposed. The first version refers to the closer-to-Sun generation scenario, suggested by \citet{jia09p05}. This version posits that the waves are produced near the sun and then transported outward by the super-Alfv\'enic solar wind. The idea tends to be reasonable based on observations of higher frequencies and larger amplitudes of LH ECWs relative to those of RH ECWs, and it is also supported by subsequent  electromagnetic simulations on the generation and propagation of ion cyclotron waves in the corona and solar wind \citep{omi14p42,omi14p50}. The other version concerns a local source characterized by unstable   particle velocity distributions. \citet{wic16p06} revealed a strong ECW storm with a duration longer than 1 hr occurring in trailing edge of the fast solar wind, and then carried out kinetic linear dispersion analyses for local proton distributions. They concluded that the storm is generated by the instability of temperature anisotropy of protons. Similar case analyses based on the local plasma parameters are also made by \citet{gar16p30} and \citet{jia16p07} but for ECWs occurring in the slow solar wind and descending part of fast solar wind, respectively. Their works demonstrate the proton velocity distributions are sufficiently anisotropic to locally drive kinetic instabilities in most of the time, six of ten intervals in \citet{jia16p07} for instance.

In particular, using the data from the \emph{STEREO} mission, \citet{zha17p08} carried out a survey of ECWs over a long period of 7 years and provided a primary indication on the mechanism of generating ECWs in the solar wind. The authors first calculated the occurrence rates of ECWs in each mouth, and found that the time-dependent occurrence rate is
nearly a constant for RH ECWs, but it varies significantly for LH ECWs. Further investigation of plasma conditions associated with occurrence of ECWs revealed that the LH ECWs take place preferentially in a plasma characterized by higher temperature, lower density, and larger velocity. Based on theoretical results concerning proton temperature anisotropy instabilities with the effect of alpha$-$proton differential flow, \citet{zha17p08} speculated that high-speed solar wind streams, and therefore alpha particle differential flow, are relevant to result in the difference of occurrence rates between LH and RH ECWs. The presence of alpha$-$proton differential flow can break the symmetry of the linear unstable waves propagating along the background magnetic field and in the opposite direction; it causes the firehose instability to preferentially generate magnetosonic waves propagating toward the Sun, and on the other hand it causes proton cyclotron instability to preferentially generate proton cyclotron waves propagating away from the Sun \citep{pod11p41}. This concept is confirmed by the most recent research based on two-dimensional hybrid simulations \citep{mar18p53}. Both instabilities will thus generate the LH ECWs in the spacecraft frame when the effect of differential flow of alpha particles relative to the protons is present.

It should be noted that direct investigation of proton temperature anisotropy was absent in the study by \citet{zha17p08}, since the \emph{STEREO} mission lacks the plasma information of proton temperature anisotropy. Based on the data from \emph{Wind} mission, this work carries out a statistical study on plasmas associated with observed ECWs, in which proton temperature anisotropy is emphasized. The paper is organized as follows. The data as well as analysis methods used in this paper are introduced in Section 2. Statistical results are presented in Section 3, and Section 4 is the summary with brief discussion.

\section{Data and analysis methods}
\emph{Wind} mission is a comprehensive solar wind laboratory in a halo orbit around the L1 Lagrange point. The magnetic field data used in this paper are obtained by the Magnetic Field Investigation (MFI) instrument sampled at a cadence of 0.092 s \citep{lep95p07}. The plasma data including ion (proton and alpha particle) perpendicular and parallel temperatures are from the Solar Wind Experiment (SWE) instrument working at a cadence of 92 s \citep{ogi95p55}. Specifically, the ion data used in this paper are produced via a nonlinear-least-squares bi-Maxwellian fit of ion spectrum from the Faraday cup \citep{kas06p05}. Note that this technique is good at ignoring proton beams, and consequently the proton data are mostly indicative of the proton core properties when there is a large proton beam \citep{kas06p05}.

Based on the high-resolution magnetic field data, we conduct a survey of ECWs occurring in the years between 2005 and 2015. To make the survey an automatic wave detection procedure is employed. The procedure was developed by \citet{zha17p79}, and improved in the other work \citep{zha18p15}. Three primary steps are carried out in the procedure. For a magnetic field interval, the first step begins with calculating the reduced magnetic helicity that is normalized and takes values in the range from $-1$ to 1 \citep[e.g.,][]{mat82p11,gar92p03,hej11p85}. The spectrum values of magnetic helicity are examined in the frequency range from 0.05 to 1 Hz. If the spectrum has absolute values $\geq$ 0.7 in some frequency band (with a minimum bandwidth of 0.05 Hz), the second step arises to identify enhanced power spectrum. The enhancement requires transverse wave power three times larger than the background power in the same frequency band; a power law fit for the entire transverse power spectrum is made to determine the background power. If the above two steps are fulfilled, the third step follows to record the wave with an amplitude criterion of 0.1 nT. During this step, a band-pass filter is used to obtain a wave amplitude \citep{wil09p06}.
The procedure can give the time intervals of ECWs occurrence, as well as their polarization senses determined directly by the sign of the spectrum values of magnetic helicity. The automation of the wave detection is achieved through dividing the long time series of magnetic field data into consecutive and overlapping time segments. Each time segment is set to be 100 s with an overlap of 80 s, implying a frequency resolution of 0.01 Hz and a time resolution of 20 s \citep{zha17p79,zha18p15}.  

\section{Statistical results}
In total 16,674,592 time segments between 2005 and 2015 are analyzed, and 339,814 (2.0\%) are identified as in the presence of ECW activities. Among the segments with ECW activities, 249,920 (74\%) segments concern LH ECWs, which is consistent with previous results \citep[e.g.,][]{boa15p10,zha18p15}. The purpose of this section is to show statistical results concerning probability density distributions (PDDs) of plasmas associated with ECW activities, occurrence rates of ECWs, and effects of alpha$-$proton differential flow. The results will be described mainly in terms of proton parallel beta ($\beta_\parallel$) and proton temperature anisotropy ($T_\perp/T_\parallel$), since $\beta_\parallel$ and $T_\perp/T_\parallel$ are two primary plasma parameters in analysis of proton temperature-anisotropy-driven instabilities \citep[i.e.,][]{gar93p67,mat13p71,wic13p03,hej18p48}. Many researches have been made by plotting data distributions in the ($\beta_\parallel$, $T_\perp/T_\parallel$) space \citep[i.e.,][]{hel06p01,bal09p01,sch11p02,mar11p01,mar12p37,hel14p15,kle18p02}.
Here the $\perp$ ($\parallel$) refers to perpendicular (parallel) with respect to the background magnetic field. 

\subsection{PDDs of ($\beta_\parallel$, $T_\perp/T_\parallel$) associated with occurrence of ECWs}
Before we move on to discuss PDDs of ($\beta_\parallel$, $T_\perp/T_\parallel$) associated with occurrence of ECWs, it is instructive to display the PDDs for ambient solar winds; in the present paper the term ``ambient" refers to all the time series irrespective of whether the ECWs are present or not. Figure 1 presents color scale plot of the ambient PDDs, where the top (bottom) panel is for the slow (fast) wind with sample number $N =$ 8,278,890 (2,846,135); the slow wind and fast wind are selected by setting the proton bulk velocity $V_p < 400$ km s$^{-1}$ and $V_p > 500$ km s$^{-1}$, respectively. One may first see that both the ambient PDDs are characterized by roughly a rhomboidal shape \citep[e.g.,][]{bal09p01,mar11p01,mar12p37,che16p26}. The second result is that the ambient PDDs depend on the solar wind types; the dominant distribution of the data points centered around ($\beta_\parallel$, $T_\perp/T_\parallel$) $\sim$ (0.71, 0.71) in case of the slow wind while it gives rise to a tendency of an anticorrelation between $\beta_\parallel$ and $T_\perp/T_\parallel$ in case of the fast wind \citep[e.g.,][]{mar04p02,hel06p01}.

Figure 2 plots the PDDs associated with wave activities in the slow wind (top panels) and the fast wind (bottom panels), where left and right panels correspond to LH and RH ECWs, respectively. In case of the slow wind, the sample number for LH ECWs is 49,004, comparable to the sample number of 41,152 for RH ECWs. In case of the fast wind, the sample number for LH ECWs is 88,787, significantly more than the sample number of 14,997 for RH ECWs. From Figure 2, one can find that the PDDs depend on the solar wind types as well as the polarization senses. The PDD for LH ECWs in the slow wind (panel (a)) is characterized by a two-population distribution, referred to as population I and population II for convenience. On the contrary, the PDD for RH ECWs (panel (b)) just shows a dispersive population and appears as a ``chunk" with probability density larger than 0.01. In case of the fast wind, the PDDs are dominated by a diagonal block structure, and the structure is clearer for LH ECWs (panel (c)).

Comparing Figure 2 with Figure 1, one can find that the PDDs associated with LH ECWs are different from the corresponding ambient PDDs. Population I in panel (a) of Figure 2 seems to be compatible with the ambient PDD shown in panel (a) of Figure 1, but the most probable values of ($\beta_\parallel$, $T_\perp/T_\parallel$) for population I is (1.12, 0.45), leading to a larger $\beta_\parallel$ as well as a smaller $T_\perp/T_\parallel$ relative to those for the ambient wind. The rise of population II is clearly not the prediction of the ambient PDD. Population II shares a positive temperature anisotropy ($T_\perp/T_\parallel > 1$) as well as a much smaller beta value of $\beta_\parallel \simeq 0.1$. Similar to the rise of population II in panel (a), there is an enhancement of probability density at region of $(\beta_\parallel$, $T_\perp/T_\parallel) \sim (0.1,3)$ in panel (c). The enhancement contributes to the clear diagonal block structure that can not be easily seen in the ambient PDD (panel (b) of Figure 1). As for the case of RH ECWs, the PDDs in principle follow the ambient PDDs and just some tiny differences between them appear; the chunk in panel (b) includes minor additional enhancement of probability density at region of $(\beta_\parallel$, $T_\perp/T_\parallel) \sim (0.1, 1)$.

\begin{figure}
\epsscale{0.9} \plotone{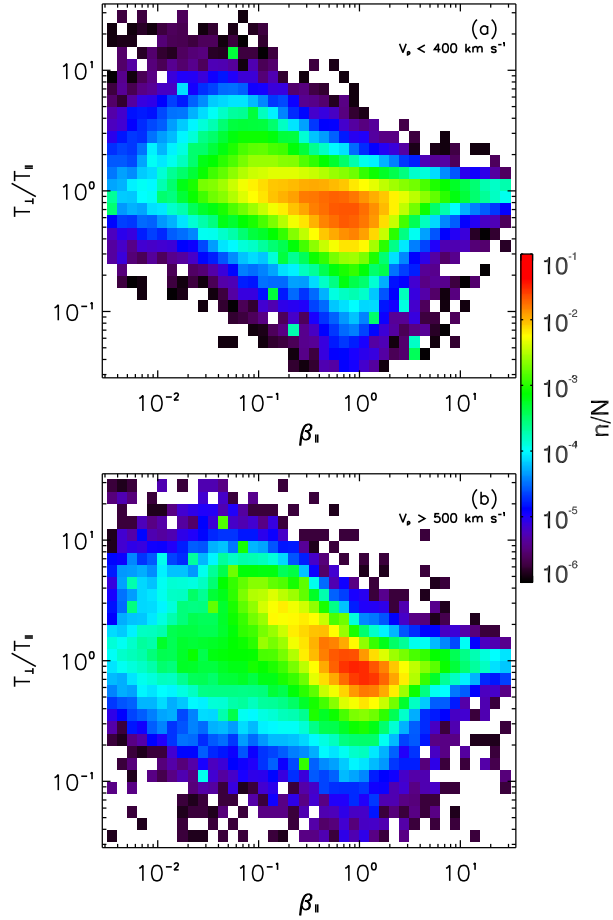} \caption{Color scale plot of PDDs of ($\beta_\parallel$, $T_\perp/T_\parallel$) for ambient solar wind observed by \emph{Wind} between 2005 and 2015, where $n$ and $N$ represent the sample number in each pixel and the total sample number in the panel, respectively. Top and bottom panels are for the slow and fast solar winds, respectively. The total sample number $N$ is 8,278,890 (2,846,135) in the top (bottom) panel.}
\end{figure}

\begin{figure}
\epsscale{0.9} \plotone{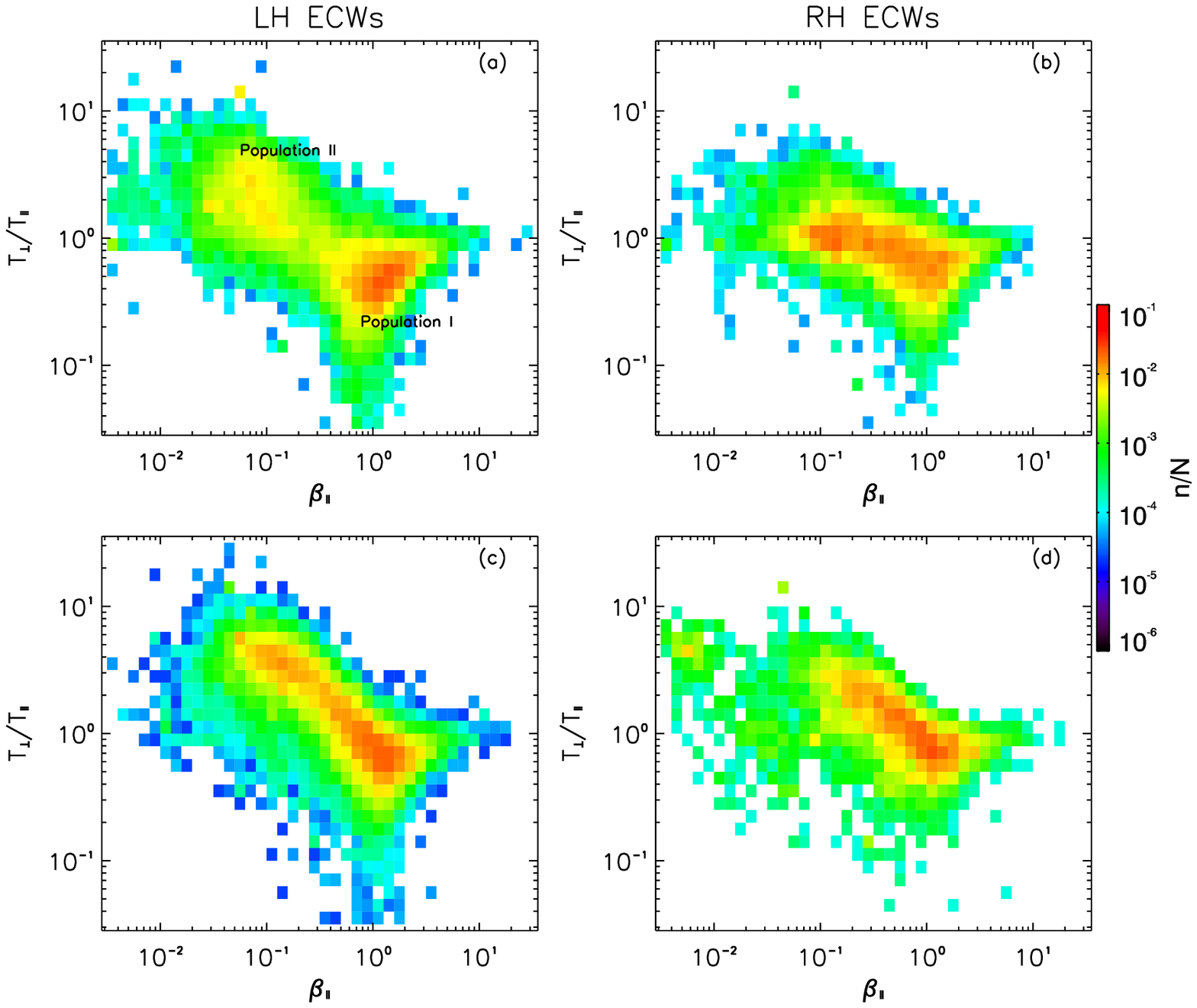} \caption{Color scale plot of PDDs of ($\beta_\parallel$, $T_\perp/T_\parallel$) associated with LH (left panels) and RH (right panels) ECWs. Top and bottom panels are for the slow and fast solar winds, respectively. The total sample numbers $N$ are 49,004, 41,152, 88,878, 14,997 in panels (a)$-$(d), respectively.}
\end{figure}

The above comparison tends to reveal preferential plasma condition for the occurrence of LH ECWs. That is a plasma with stronger proton temperature anisotropies. Moreover, the preferential condition requires a very small beta value of $\beta_\parallel \simeq 0.1$ if the plasma is with positive temperature anisotropy, or a larger beta value of $\beta_\parallel \gtrsim 1$ if it is with negative temperature anisotropy. The preferential plasma condition for the occurrence of RH ECWs seems to be absent.

\subsection{Occurrence rates of ECWs and temperature-anisotropy-driven instabilities}
Overall, the occurrence rates of ECWs are very different for different solar wind types. The occurrence rate is about 1.1\% in the slow solar wind, while it is up to 3.6\% in the fast solar wind. In the ($\beta_\parallel$, $T_\perp/T_\parallel$) space, nonuniform occurrence rates of ECWs can be expected since the PDDs associated with ECWs do not match with the ambient PDDs. Figure 3 plots the occurrence rates calculated according to sample numbers of ECWs and ambient plasmas; for a calculation we require the sample number of ambient plasmas exceeding 500 in each pixel of ($\beta_\parallel$, $T_\perp/T_\parallel$). Similar to the format in Figure 2, left (right) panels in Figure 3 are for LH (RH) ECWs, and top (bottom) panels refer to the slow (fast) solar wind. The result first shows that the plasmas with proton temperature considerably departing from isotropy can lead to rise of ECWs. For $T_\perp/T_\parallel > 3$ or $T_\perp/T_\parallel < 0.3$, the occurrence rate of ECWs can exceed 10\% or be up to 25\% that is much larger than the average level. For LH ECWs, there is a tendency that the occurrence rate increases as the {proton temperature anisotropy increases. In particular, the betas ($\beta_\parallel$) with maximum of occurrence rates are significantly different for different temperature anisotropies; it is near 0.1 when $T_\perp/T_\parallel > 1$ while it is around 1 when $T_\perp/T_\parallel < 1$.

\begin{figure}
\epsscale{0.9} \plotone{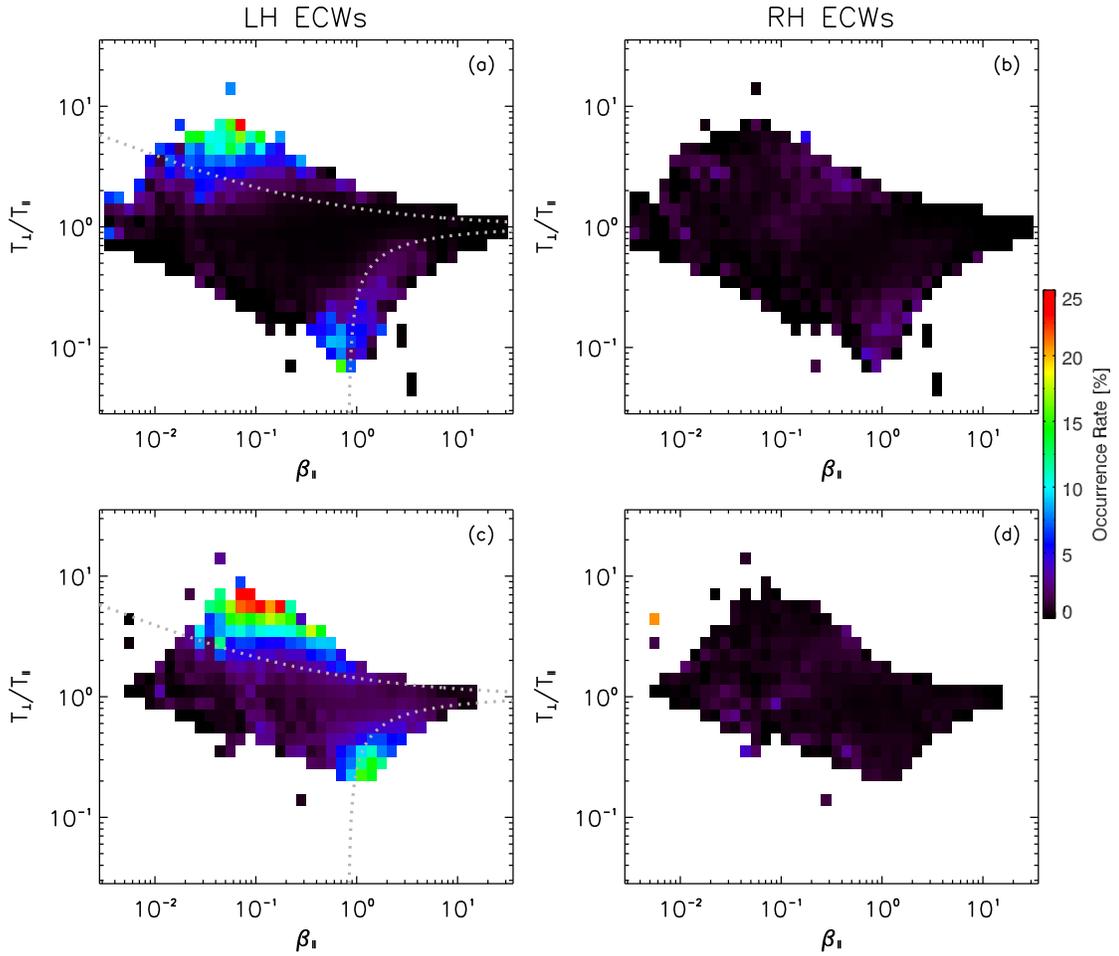} \caption{Color scale plot of occurrence rates of ECWs. Left and right panels are for LH and RH ECWs, while top and bottom panels are for the slow and fast solar winds, respectively. The grey dotted lines in left panels are for thresholds of proton cyclotron instability (upper lines) and parallel firehose instability \citep[lower lines;][]{hel06p01}.}
\end{figure}

It is well known that a plasma with proton temperature $T_{\perp} > T_\parallel$ can excite proton cyclotron waves by proton cyclotron instability, while a plasma with $T_{\perp} < T_\parallel$ may generate magnetosonic waves by parallel firehose instability. Both instabilities should be relevant for the present study. The observed ECWs in the present paper are characterized by frequencies near the proton cyclotron frequency and by quasi-parallel (or antiparallel) propagation, which is in line with the theoretical prediction.
On the other hand, kinetic theory also predicts some threshold conditions with appreciable growth of the instabilities in terms of ($\beta_\parallel$, $T_\perp/T_\parallel$). The conditions corresponding a growth rate of $10^{-3}\omega_{cp}$ are plotted as grey dotted lines in left panels of Figure 3, where $\omega_{cp}$ is the proton cyclotron frequency \citep{hel06p01}. The present result, in particular, shows that plasmas with satisfaction of the threshold conditions often contribute to a much large occurrence rate of (LH) ECWs, especially in the case of the fast wind (panel (c)). The absence of large occurrence rate with larger $\beta_\parallel$ might be due to the competition from other kinetic instabilities, such as the mirror and oblique firehose instabilities that produce waves with properties distinctly different from those of the present ECWs \citep[e.g.,][]{taj67p82,gar92p19,hel00p19}. In addition, one may note that there are still wild occurrence of ECWs (with occurrence rates $\gtrsim 5\%$ in left panels) in the regions where $\beta_\parallel$ and $T_\perp/T_\parallel$ approach but do not satisfy the threshold conditions marked by the grey dotted lines. One reason may be the presence of alpha$-$proton differential flow, which will be discussed in the next subsection.

In the case of RH ECWs, the result is complicated a bit. As shown in panel (b), it is not so clear for the tendency of occurrence rate increasing with proton temperature anisotropy. In panel (d), this tendency can not be found, since some large occurrence rates around 2\% appear even though the proton temperatures are isotropic. Of course, one should note that the occurrence rate for RH ECWs is usually much low relative to that for LH ECWs.

\subsection{Effects of alpha$-$proton differential flow}
The ion data described in Section 2 allow us to investigate the parameters for alpha particles, including the density, differential velocity, and temperature anisotropy. Results concerning the temperature anisotropy seem to be complicated and irregular, while those related to the density, differential velocity are relevant. To illustrate the relevance Figures 4 and 5 plot PDDs of ($V_d/V_A$, $N_{\alpha}/N_{p}$) for ambient solar winds and those associated with ECWs with formats similar to those in Figures 1 and 2, respectively, where $V_d$ ($V_A$) is the differential velocity with respect to protons (local Alfv\'en velocity), and $N_{\alpha}$ ($N_{p}$) is the number density of alpha particle (proton). Figure 4 shows that the ambient PDDs depend on the solar wind types and the fast wind shares a larger differential velocity as well as a higher density of alpha particle; the  $V_d/V_A$ ($N_{\alpha}/N_{p}$) is often less than 0.2 (0.03) in the slow wind while it is frequently greater than 0.4 (0.04) in the fast wind. For the PDDs associated with ECWs (Figure 5), they depend on not only the wind types but also the wave polarizations. The PDDs for LH ECWs have enhanced probability densities at regions with larger $V_d/V_A$ and/or higher $N_{\alpha}/N_{p}$ relative to the PDDs for RH ECWs. A similar difference appears by comparing the PDDs for LH ECWs with the corresponding ambient PDDs, contributing to nonuniform wave occurrence rates. The distributions of wave occurrence rates in ($V_d/V_A$, $N_{\alpha}/N_{p}$) space, plotted in Figure 6, indicate large occurrence rates arising mainly at regions with larger $V_d/V_A$ and/or higher $N_{\alpha}/N_{p}$ for LH ECWs (Left panels). These results tend to suggest that LH ECWs occur preferentially in plasmas with larger differential velocity and/or higher alpha particle density.

We note that usually only the parameter $V_d/V_A$ is investigated in existing literatures in which a fixed $N_{\alpha}/N_{p}$ is used \citep[e.g.,][]{hel06p07,pod11p41}. The present data show that both $V_d/V_A$ and $N_{\alpha}/N_{p}$ are important to the occurrence of ECWs; according to Figure 6, the $N_{\alpha}/N_{p}$ seems to be a more sensitive parameter for determining the wave occurrence rates. In this regard, an integrated parameter including $N_{\alpha}/N_{p}$ should be helpful to discuss effects of alpha$-$proton differential flow. We thus employ the normalized kinetic energy defined by $\xi_{\alpha} = m_{\alpha}N_{\alpha}V_d^2/m_{p}N_{p}V_A^2$, where $m_{\alpha}$ ($m_{p}$) is the mass of alpha particle (proton). It is found that the occurrence of ECWs has a significant dependence on the normalized kinetic energy. Figure 7 displays the dependence of occurrence rates of ECWs on the kinetic energy $\xi_{\alpha}$ less than 0.125; above which the occurrence rates fluctuant randomly (not shown). These surveys are made in different bins with observations exceeding 5000 in each bin. In Figure 7, similar to the format of Figures 2 and 3, left (right) panels are for LH (RH) ECWs, and top (bottom) panels refer to the slow (fast) solar wind. One can see that the dependence is different for different solar wind types as well as wave polarizations. In case of the slow wind, the occurrence rates of LH and RH ECWs are comparable when $\xi_{\alpha}$ is very small. As $\xi_{\alpha}$ increases from approximately zero to 0.07, the occurrence rate of LH ECWs increases quickly from about 0.3\% to 1.8\% and becomes fluctuant when $\xi_{\alpha}$ is greater than 0.07. For the RH ECWs (panel (b)), although the occurrence rate rises from about 0.4\% to 0.9\% when $\xi_{\alpha}$ goes to 0.04, it does not vary much relative to the case of LH ECWs. In the fast wind, the occurrence rate of LH ECWs is mainly around 2.7\% when $\xi_{\alpha} \lesssim 0.08$ while it rises rapidly up to about 4.4\% as $\xi_{\alpha}$ increases. For the RH ECWs (panel (d)), their occurrence rate seems to decrease with $\xi_{\alpha}$ and is always much low relative to that of LH ECWs in the fast wind.
Overall, a larger alpha$-$proton differential kinetic energy $\xi_{\alpha}$ contributes to a higher occurrence rates of ECWs when $\xi_{\alpha} \lesssim 0.125$.

\begin{figure}
\epsscale{0.9} \plotone{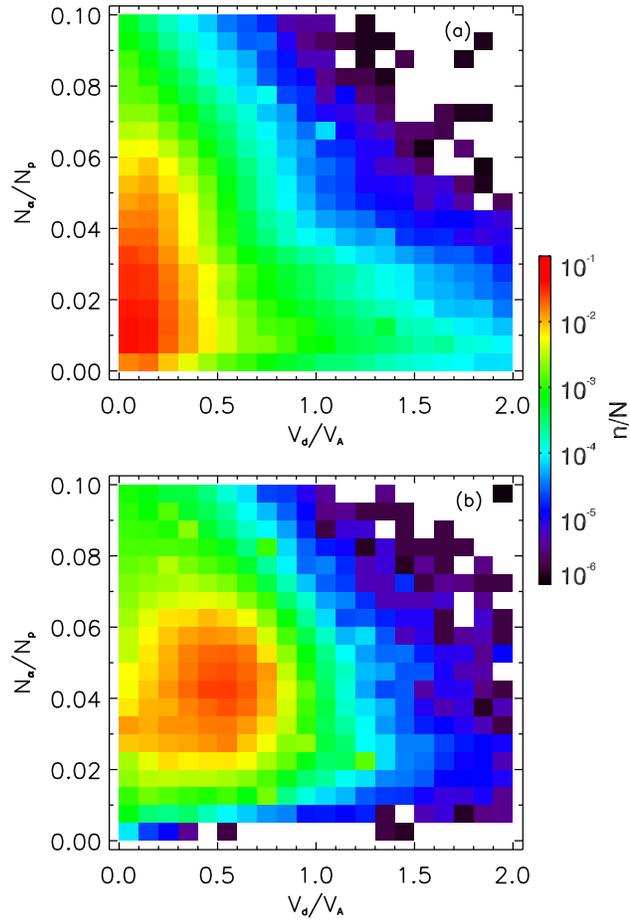} \caption{Color scale plot of PDDs of ($V_d/V_A$, $N_{\alpha}/N_{p}$) for ambient solar wind observed by \emph{Wind} between 2005 and 2015. Top and bottom panels are for the slow and fast solar winds, respectively.}
\end{figure}
\begin{figure}
\epsscale{0.9} \plotone{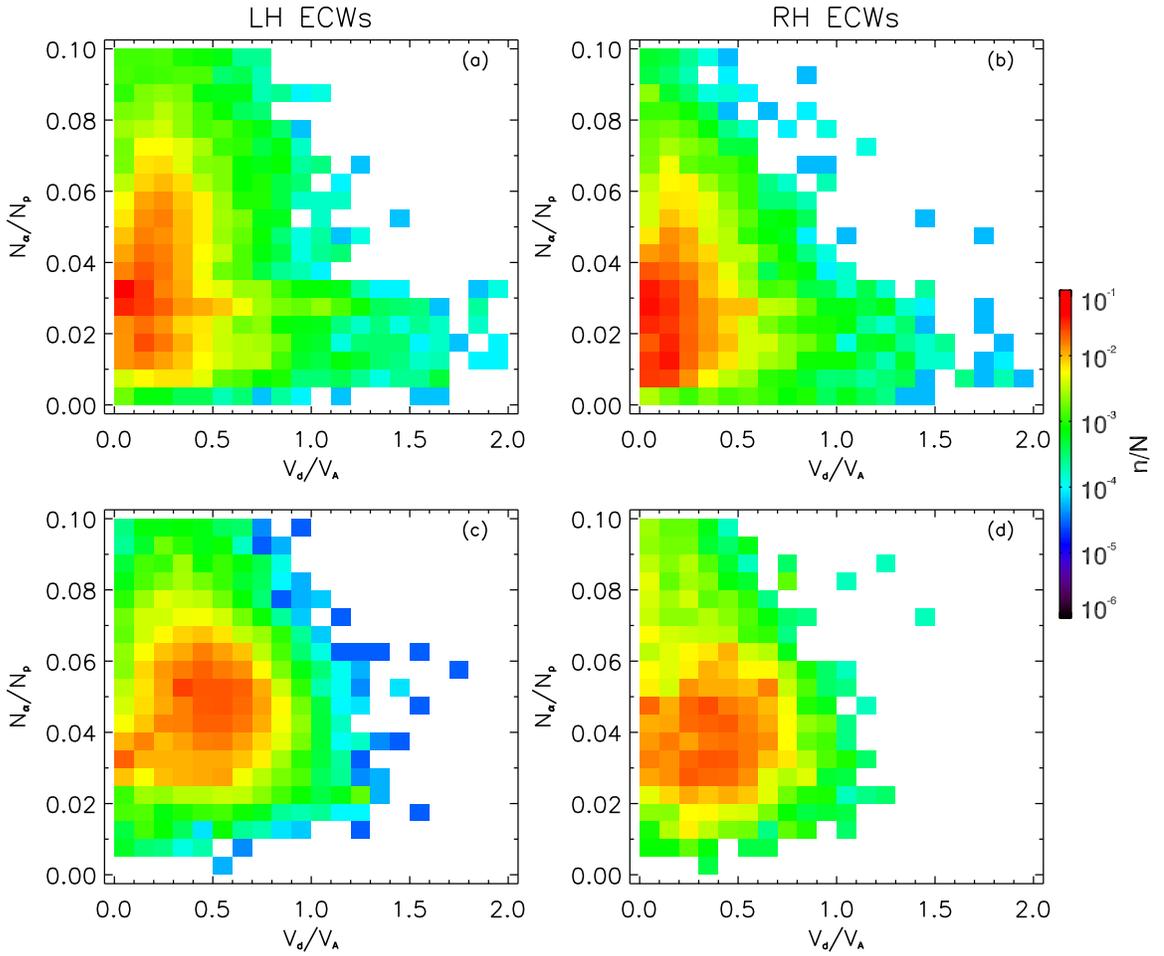} \caption{Color scale plot of PDDs of ($V_d/V_A$, $N_{\alpha}/N_{p}$) associated with LH (left panels) and RH (right panels) ECWs. Top and bottom panels are for the slow and fast solar winds, respectively. }
\end{figure}

\begin{figure}
\epsscale{0.9} \plotone{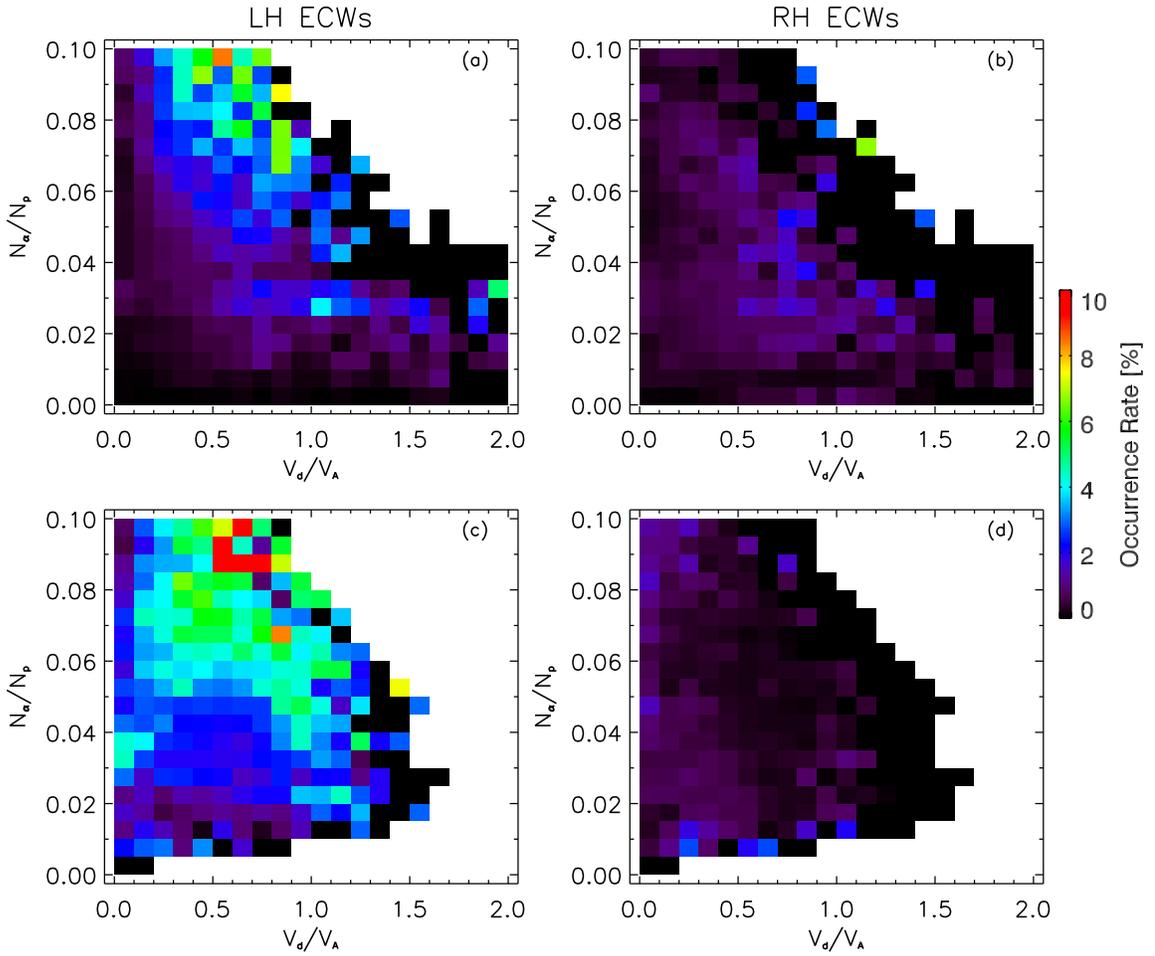} \caption{Color scale plot of occurrence rates of ECWs in the ($V_d/V_A$, $N_{\alpha}/N_{p}$) space. Left and right panels are for LH and RH ECWs, while top and bottom panels are for the slow and fast solar winds, respectively.}
\end{figure}

\begin{figure}
\epsscale{0.9} \plotone{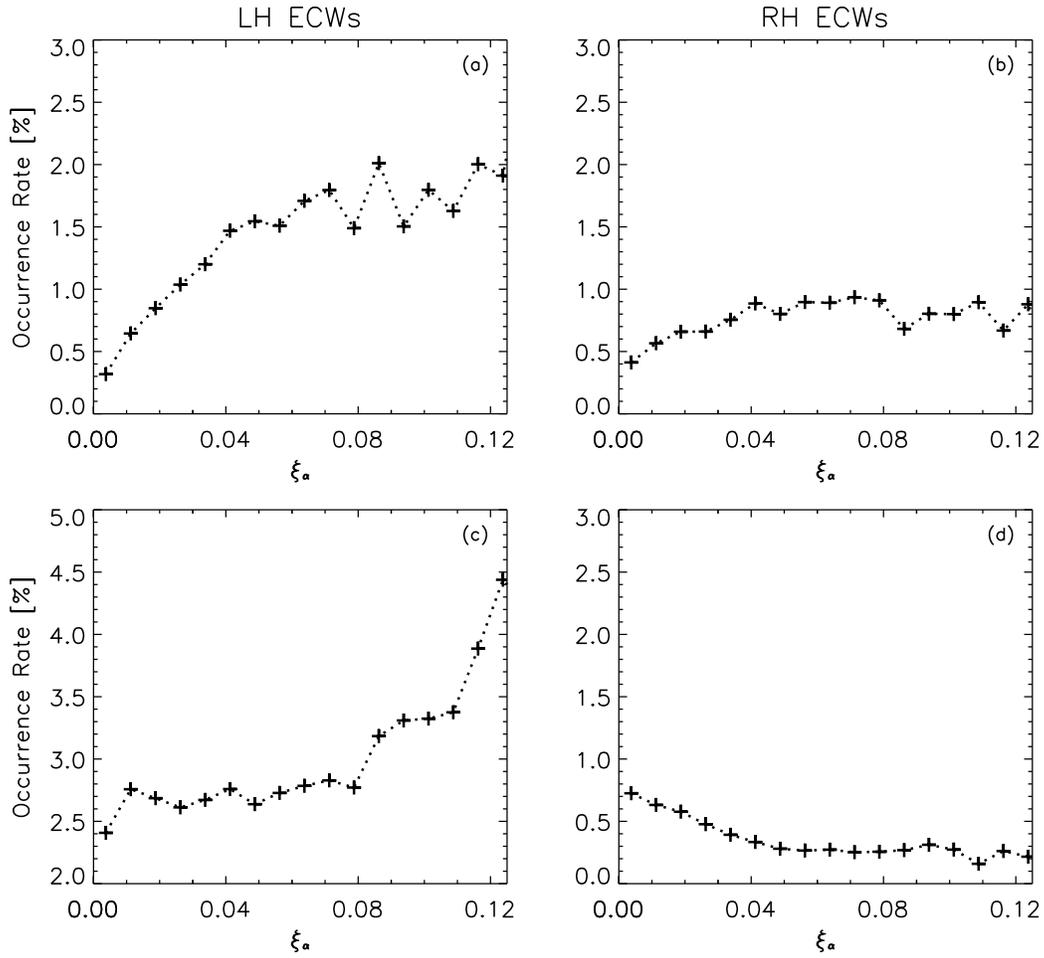} \caption{Occurrence rates of ECWs against normalized alpha$-$proton differential kinetic energy ($\xi_{\alpha}$). Left and right panels are for LH and RH ECWs, while top and bottom panels are for the slow and fast solar winds, respectively.}
\end{figure}
\begin{figure}
\epsscale{0.9} \plotone{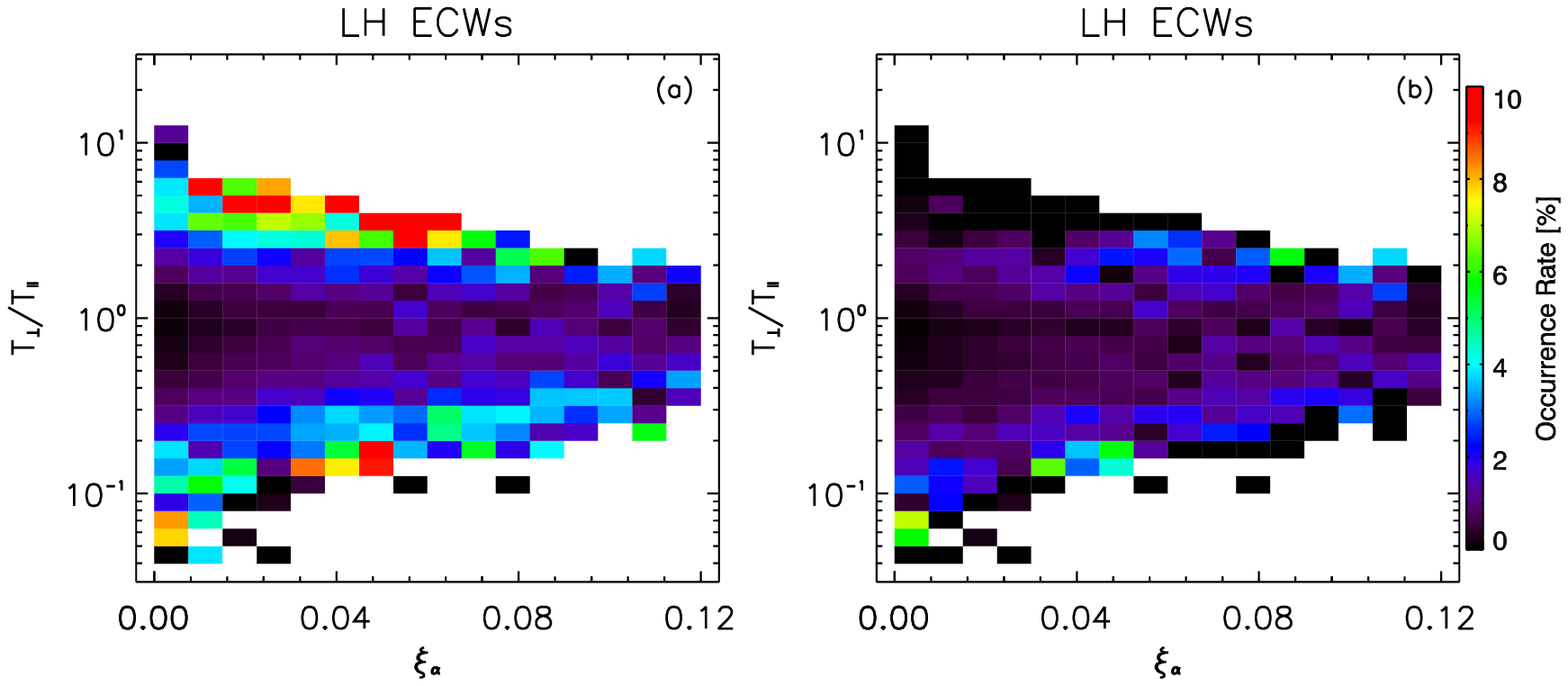} \caption{Color scale plot of occurrence rates of LH ECWs in the slow wind against normalized alpha$-$proton differential kinetic energy ($\xi_{\alpha}$) and proton temperature anisotropy ($T_\perp/T_\parallel$). All ECWs are collected in panel (a) while just the ECWs with parameters dissatisfying the instability threshold conditions (grey dotted lines in Figure 3) are counted in panel (b).}
\end{figure}

On the other hand, the presence of alpha$-$proton differential flow with large kinetic energy seems to can lower the threshold condition for excitation of the temperature-anisotropy-driven instabilities. For illustrating this point Figure 8 is presented which is for LH ECWs in the slow wind. From this figure, large occurrence rates mainly arise with a distinct temperature anisotropy and/or large differential kinetic energy. Much low occurrence rates ($< 0.2\%$) appear at $T_\perp/T_\parallel \sim 1$ when $\xi_{\alpha}$ is very small ($<$ 0.01). However, if $\xi_{\alpha} \gtrsim 0.06$ is fulfilled, occurrence rates can be considerably high ($\gtrsim 2\%$) even though $T_\perp/T_\parallel$ approaches unity.
It is likely because the presence of alpha$-$proton differential flow results in greater growth rates of instabilities, or equivalently lower the threshold condition for excitation of the instabilities in terms of ($\beta_\parallel$, $T_\perp/T_\parallel$) since the threshold condition is determined by a given growth rate, e.g, $10^{-3}\omega_{cp}$ \citep{hel06p01}. To test this idea, panel (b) is plotted where the data are the same as those in panel (a), but just count the ECWs with parameters dissatisfying the traditional threshold condition represented be grey dotted lines in Figure 3. It is interesting that the profiles of regions with $T_\perp/T_\parallel$ around unity in both panels are comparable.

According to our ECW sample, a relevant issue is that the percentages of LH ECWs are significantly different between the cases of slow and fast winds; it is about 54\% in case of the slow wind while it is up to 86\% in case of the fast wind. This may reinforce the interpretation in terms of the effect of alpha$-$proton differential flow since the differential flow prevails in the fast wind as shown in Figures 4 and 5. From Figure 4, the most probable values of ($V_d/V_A$, $N_{\alpha}/N_{p}$) in the fast wind are about (0.55, 0.043), which are much greater than those in the slow wind, i.e., (0.05, 0.008). According to the theory by \citet[][their Figure 3]{pod11p41}, the larger differential velocity ($V_d/V_A$), and therefore larger differential kinetic energy, makes magnetosonic waves propagating toward the Sun stronger, but reduces the growths of magnetosonic waves propagating away from the Sun. Consequently, the larger differential kinetic energy results in less presence of RH ECWs with a given amplitude criterion and therefore contributes to a larger percentage of LH ECWs in the spacecraft frame. 

\section{Summary and discussion}
In this paper, we investigate the ion data set from Wind/SWE instrument with simultaneous detection of ECWs during the years 2005$-$2015. We first make a comparison between PDDs of ($\beta_\parallel$, $T_\perp/T_\parallel$) for ambient solar winds and for the solar winds with occurrence of ECWs. The PDDs with LH ECWs are different from the ambient PDDs, and the former have probability density enhancements at regions with $\beta_\parallel \sim 0.1$ and $T_\perp/T_\parallel > 1$. Consequently, a two-population distribution and a diagonal block structure appear in the PDDs with LH ECWs for the slow and fast solar winds, respectively. Distributions of ECW occurrence rates are obtained in the ($\beta_\parallel$, $T_\perp/T_\parallel$) space. The distributions for LH ECWs reveal the trend of occurrence rate increasing with proton temperature anisotropy either for the slow wind or for the fast wind. Further, a much large occurrence rate often arises when plasmas with parameters satisfying threshold conditions for excitation of proton temperature-anisotropy-driven instabilities. Based on investigation of alpha particle parameters, it is found that the normalized kinetic energy of alpha$-$proton differential flow is a better parameter to describe the occurrence of LH ECWs. Double effects of the differential flow are implied. One is the contribution of a larger occurrence rate of ECWs, and the other is for a higher percentage of LH ECWs.

The present results may provide an indication on the mechanism of generating LH ECWs. Results from Figure 3 show that the plasmas with proton temperature considerably departing from isotropy can lead to larger occurrence rates of LH ECWs. Moreover, the $\beta_\parallel$ with maximum of occurrence rates are different for different temperature anisotropies; it is near 0.1 when $T_\perp/T_\parallel > 1$ and it is around 1 when $T_\perp/T_\parallel < 1$. These results are in line with the theory for proton cyclotron and parallel firehose instabilities \citep{gar93,Gar15p49,yoo17p04}. We thus speculate both instabilities are relevant to generate these ECWs. The results concerning effects of alpha$-$proton differential flow reinforce this speculation. Existing theory and simulation show that the presence of alpha$-$proton differential flow contributes to a larger growth rate of the instabilities and meanwhile results in the LH polarization of ECWs in the solar wind \citep{hel06p07,pod11p41}. Finally, we propose that the proton cyclotron and parallel firehose instabilities with effects of alpha$-$proton differential flow are two mechanisms of locally generating ECWs in the solar wind.

As for RH ECWs, their generation mechanism seems to be complicated. Their behaviors are significantly different from those of LH ECWs. Previous results have revealed that their monthly occurrence rate is approximately a constant, i.e., not showing great fluctuation as that for LH ECWs \citep{zha17p08}. The preferential plasma conditions for the wave activities as well as the tendency of occurrence rate increasing with the proton temperature anisotropy can not be found (at least not clear). Relative to LH ECWs, they tend to be pervasive in all region of ($\beta_\parallel$, $T_\perp/T_\parallel$), by comparing panel (d) with panel (c) in Figure 2. Previous results also showed that the properties of RH ECWs (frequency, amplitude, and normal angle) are characterized by more dispersive distributions than those for LH ECWs \citep{zha18p15}. These results may imply complex processes for generation of RH ECWs. A specific mechanism can not be obtained in this paper but two comments may be helpful. First, a very slow solar wind may contribute to the presence of RH ECWs and produce enhanced probability density at region of $(\beta_\parallel$, $T_\perp/T_\parallel) \sim (0.1,1)$ as shown in panel (b) of Figure 2. Our primary test in terms of solar wind velocity distribution in the space of ($\beta_\parallel$, $T_\perp/T_\parallel$) shows that the region of $(\beta_\parallel$, $T_\perp/T_\parallel) \sim (0.1,1)$ corresponds to very slow solar winds with median velocity ($\sim$350 km s$^{-1}$) that is smaller than those in other regions. Such slow winds perhaps allow ECWs propagate far enough away from their source regions and become nonlocal ECWs when they are observed. Second, the result from panel (d) in Figure \textbf{5} tends to imply that the generation mechanism for RH ECWs in the fast wind is not likely to be the proton cyclotron or parallel firehose instability with effects of alpha$-$proton differential flow since the flow would lead to LH ECWs whereas observations appear still as RH ECWs.
In this sense, nonlocal processes or other generation mechanisms need to be examined. The first mechanism, for instance, maybe the proton/proton magnetosonic instability driven by a relatively cool proton beam streaming fast enough with differential velocity typically greater than the local Alfv\'en velocity \citep[e.g.,][]{mon75p67,mon76p43,dau98p13,dau99p57}. The second mechanism, similar to the case of proton beam, is perhaps the alpha/proton magnetosonic instability when a fast alpha particle beam moving with respect to core protons is in the presence \citep[e.g.,][]{gar00p55,lix00p83,luq06p01,ver13p88}. Note that magnetosonic waves produced by the above two instabilities could be observed mainly as RH ECWs, because these magnetosonic waves should propagate mainly away from the Sun and the majority of the waves do not suffer polarization reversal \citep{gar93,luq06p01}. (Explicitly speaking, the waves propagate in the direction of the beam that usually points away from the Sun.) Another possible mechanism to produce RH
ECWs is the parallel firehose instability of alpha particles \citep{mar12p37,ver13p63}. Recent studies show that magnetosonic waves produced by this instability propagate preferentially in the direction of the alpha particle beam \citep{mat15p13,seo16p13}.

Before concluding, two remarks on the present study may be appropriate. First, the present study just refers to the properties of proton cores and alpha particles. A comprehensive research including proton beam parameters should be made in the future since proton beams are common in the fast solar wind and can also serve as free-energy sources for instabilities \citep[e.g.,][]{mar87p63,mat13p71,kle18p02}. Second, we emphasize the solar wind type is determined actually by the solar wind velocity in the present paper. One should keep in mind that the solar wind velocity is not necessarily a good parameter for characterization of the solar wind. Some solar wind streams with velocities less than 400 km s$^{-1}$ have many properties (e.g., high degree of Alfv\'enicity) similar to standard fast solar wind streams \citep{ami16p02}. Further investigation of the present study with distinguishing solar winds in terms of physical properties or their source regions on the solar surface (coronal streamers, active regions, or coronal holes) is desirable.

\acknowledgments
The authors thank NASA/GSFC for the use of data from the \emph{Wind} mission, which are available freely via the Coordinated Data Analysis Web (http://cdaweb.gsfc.nasa.gov/cdaweb/istp$_-$public/). This research was supported by NSFC under grant Nos. 41504131, 41874204, 41674170, 41531071, and Scientific Research Projects in Universities of Henan Province (19HASTIT020, 17A170002). This research was also sponsored partly by the Key Laboratory of Solar Activity at CAS NAO (KLSA201703). The authors are grateful to the anonymous referee for valuable comments to improve this paper.


\end{document}